%% file: Proc_TOP2021.tex
\documentclass[12pt]{article}
\usepackage{graphicx}
\usepackage{amssymb}

\newcommand\pubnumber{DESY 22-002}
\newcommand\pubdate{\today}

\def\institute{$^\mathrm{a}\,$II. Institut f\"ur Theoretische Physik, Universit\"at Hamburg, \\ Luruper Chaussee 149, D~--~22761 Hamburg, Germany\\
$^{\mathrm{b}}\,$Center For Particle Physics Siegen, Department Physik,  Universit\"at Siegen, \\
  Walter Flex Str. 3, D~--~57068 Siegen, Germany}
\def\support{}
\def\Title#1{\begin{center} {\Large #1 } \end{center}}
\def\Author#1{\begin{center}{ \sc #1} \end{center}}
\def\Address#1{\begin{center}{ \it #1} \end{center}}

\newcommand\pubblock{\rightline{\begin{tabular}{l} \pubnumber\\
         \pubdate  \end{tabular}}}
\newenvironment{Abstract}{\begin{quotation}  }{\end{quotation}}
\newenvironment{Presented}{\begin{quotation} \begin{center} 
             PRESENTED AT\end{center}\bigskip 
      \begin{center}\begin{large}}{\end{large}\end{center} \end{quotation}}

\input econfmacros.tex
\begin{document}
\begin{titlepage}
\pubblock

\vfill
\Title{Uncertainties on the theoretical input for $t\bar{t}j$ experimental analyses at the LHC}
\vfill
\Author{ Katharina Vo{\ss}$^{\mathrm{a},\mathrm{b}}$, Maria Vittoria Garzelli$^{\mathrm{a}}$, Sven-Olaf Moch$^{\mathrm{a}}$\support}
\Address{\institute}
\vfill
\begin{Abstract}
The precise measurement of the top-quark mass constitutes one of the main goals of the LHC top-quark physics program. One possibility to extract this parameter uses the $\rho_{\mathrm{s}}$ distribution, which depends on the invariant mass of the $t\bar{t}j$ system. To fully take advantage of the experimental accuracy achievable in measuring top quark production cross sections at the LHC, the theory uncertainties need to be well under control. We present a study of the effect of varying the input parameters of the theoretical calculation on the predicted $\rho_{\mathrm{s}}$ distribution. Thereby we studied the influence of the $R$ parameter in the jet reconstruction procedure, as well as the effect of various renormalization and factorization scale definitions and different PDF sets. A behaviour similar to the one presented here for the $\rho_{\mathrm{s}}$-distribution was also found for other differential distributions.
\end{Abstract}
\vfill
\begin{Presented}
$14^\mathrm{th}$ International Workshop on Top Quark Physics\\
(videoconference), 13--17 September, 2021
\end{Presented}
\vfill
\end{titlepage}
\def\thefootnote{\fnsymbol{footnote}}
\setcounter{footnote}{0}

\section{Introduction}
A possibility to extract the top-quark mass $m_t$ in a well defined mass renormalization scheme is given by the analysis of the normalized $\rho_{\mathrm{s}}$ distribution, as first discussed in~\cite{Alioli:2013}, which is defined as $(1/{\sigma_{t\bar{t}j}}) \, {d \sigma_{t\bar{t}j}}/{d \rho_\mathrm{s}}$, with $\rho_s = {2 m_0}/{m_{t\bar{t}j}}$ and $m_0 = 170\,\mathrm{GeV}.$ \par 
A recent extraction of the ATLAS collaboration \cite{ATLAS:2019JHEP} of $m_t$ in $t\bar{t}j$ hadroproduction using the $\rho_{\mathrm{s}}$ distribution led to an on-shell value of $m_t^{\mathrm{pole}} = 171.1 \pm 0.4 (\mathrm{st}) \pm 0.9 (\mathrm{sy})\, ^{+0.7}_{-0.3}(\mathrm{th})\,\mathrm{GeV},$
in which the theory uncertainty is sizeable. Thereby, the scale uncertainty dominates the theory uncertainty and contributes to an uncertainty on $m_t^{\mathrm{pole}}$ with $^{+0.6}_{-0.2}\,$GeV, while the PDF and $\alpha_S$ uncertainty amounts to $\pm 0.2\,$GeV.\par 
In \cite{Bevilacqua:2016} a better description of the high energy tails of differential NLO distributions of the process $pp \rightarrow t\bar{t}j$ with fully leptonic top-quark decay was found when using dynamical renormalization and factorization scales ($\mu_R,\mu_F$) instead of a static scale $\mu_0=m_t$. We therefore present a study of the scale and parton distribution function (PDF) uncertainty using a static and two dynamical scales, with more inclusive cuts in contrast to \cite{Bevilacqua:2016}, similar to those recently adopted by the experimental collaborations, and a special focus on the discussed $\rho_{\mathrm{s}}$ distribution.

\section{Theory uncertainty in the $\rho_{\mathrm{s}}$ distribution}
\begin{figure}[!h!tbp]
\centering
\includegraphics[width=\textwidth]{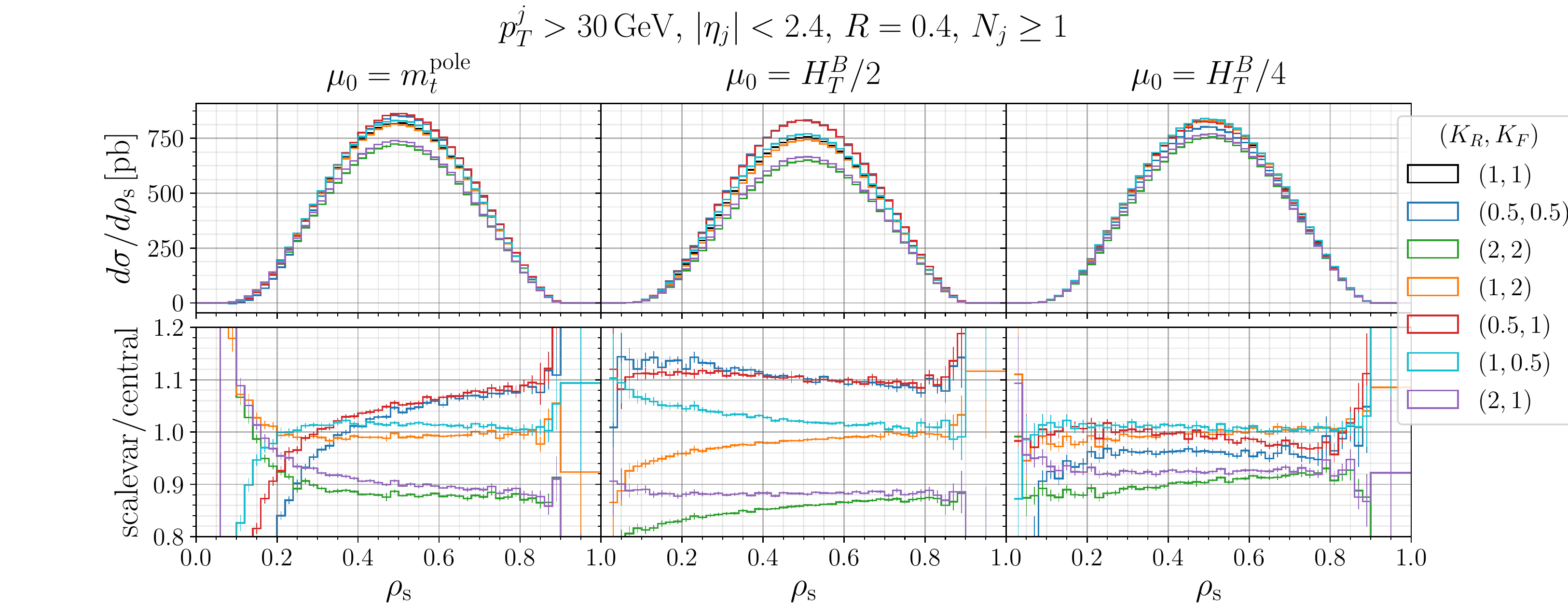}
\caption{NLO prediction of the $\rho_\mathrm{s}$ distribution and the seven point scale variation using as nominal $\mu_R,\mu_F$ the scales $\mu_0=m_t,\, H_T^B/2$ and $H_T^B/4$ (from left to right).}
\label{fig:scalevar_3panels}
\end{figure}

\begin{figure}[!h!tbp]
\centering
\includegraphics[width=\textwidth]{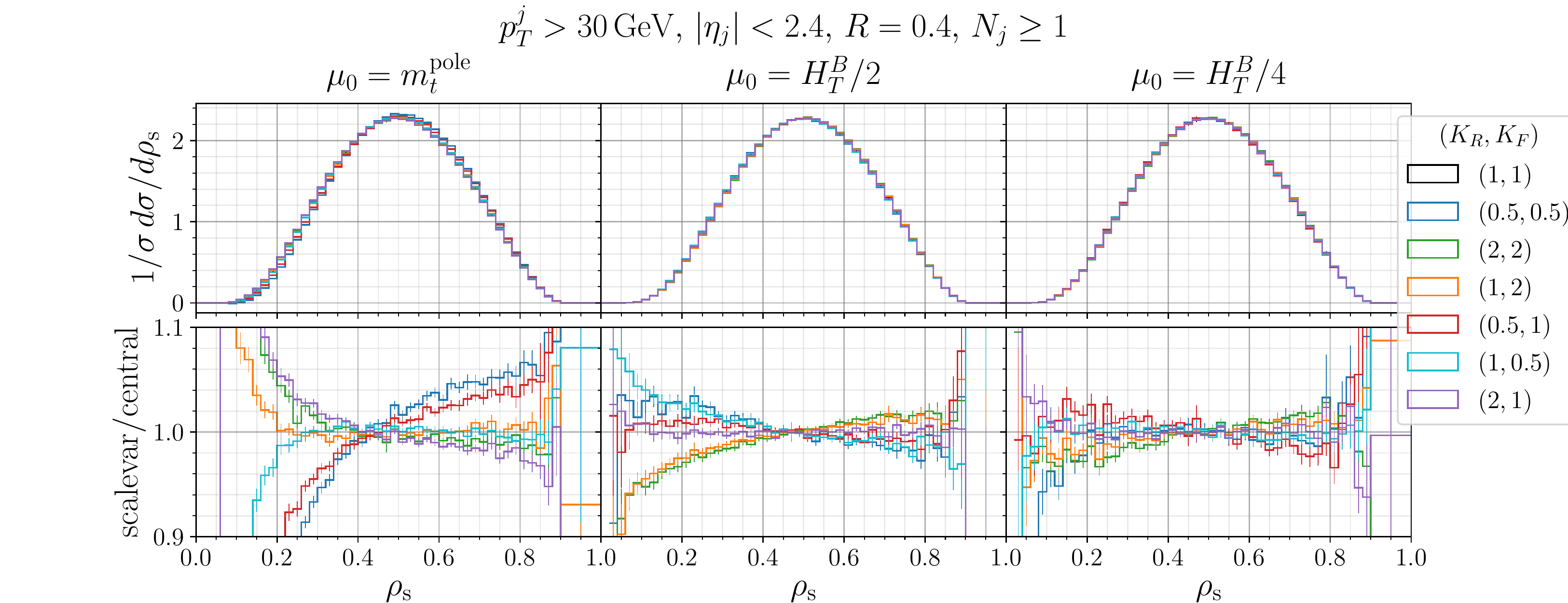}
\caption{Same as in Fig.~\ref{fig:scalevar_3panels}, but for the normalized $\rho_\mathrm{s}$ distribution.}
\label{fig:scalevar_3panels_norm}
\end{figure}

The NLO differential cross sections of the $pp \rightarrow t\bar{t}j$ process at $\sqrt{s}=13\,$TeV for $m_t^{\mathrm{pole}}=172\,$GeV presented in the following were obtained with the \texttt{ttbarj V2} implementation in the POWHEG-BOX \cite{Alioli:2010}. In contrast to the previous \texttt{ttbarj V1} implementation \cite{Alioli:2012}, in \texttt{ttbarj V2} all hard scattering amplitudes are calculated with \texttt{OpenLoops2} and the calculation can be parallelized. During the analysis at least one jet satisfying the kinematic cuts $p_T^j > 30\,$GeV and $|\eta_j|<2.4$ was required, with jets reconstructed using the anti-$k_T$ jet clustering algorithm with $R=0.4$. The default PDFs and $\alpha_S(M_Z)$ value were taken from the CT18NLO PDF set. Besides the static scale $\mu_0=m_t^{\mathrm{pole}}$ also the two dynamical scales $\mu_0=H_T^B/2$ and $H_T^B/4$ were investigated defined through $H_T^B = \sqrt{{p_{T,t}^B}^2 + {(m_t^{\mathrm{pole}})}^2} + \sqrt{{p_{T,\bar{t}}^B}^2 + {(m_t^{\mathrm{pole}})}^2} + p_{T,j}^B$. The superscript $B$ means that the kinematic variables are evaluated at the underlying Born level in the POWHEG-BOX.\\
In Fig.~\ref{fig:scalevar_3panels} the $\rho_{\mathrm{s}}$ distribution and the corresponding seven point scale variation graphs, using the three scale definitions described above, are shown explicitly. The graphs are obtained from $(\mu_R,\mu_F) = (K_R,K_F)\mu_0$ by varying $K_R, K_F \in \{ 0.5, 1, 2\}$, leaving out the extreme combinations $(0.5,2)$ and $(2,0.5)$. In fact, in case of the static scale, for low values of $\rho_{\mathrm{s}}$, which correspond to large values of $m_{t\bar{t}j}$ and as such to the high-energy region, the description of the $\rho_s$ distribution seems to be unreliable and the scale uncertainty increases rapidly. Further, a crossing of the graphs obtained with different values of $K_R, K_F$ is seen in the interval $0.1 \lesssim \rho_{\mathrm{s}} \lesssim 0.3$. These distinctive features are not observed using either dynamical scale and the scale variation does not induce such a large shape variation as seen in the static scale case.
This leads to the observation that the scale uncertainty in the normalized $\rho_{\mathrm{s}}$ distribution is clearly reduced using a dynamical instead of the static scale, as shown in Fig.~\ref{fig:scalevar_3panels_norm}. 
Further investigation of the scale uncertainty showed that using a dynamical scale with either $R=0.4$ or $R=0.8$ leads to a similar scale uncertainty, while in the static scale case the scale uncertainty can be reduced in the region of low $\rho_{\mathrm{s}}$ using $R=0.8$. Comparing the NLO and LO scale variation uncertainty bands, these overlap over the whole region of the $\rho_{\mathrm{s}}$ distribution using either dynamical scale, while in the predictions obtained with the static scale, the NLO and LO scale variation bands start to depart from each other in the high-energy region.\\
Furthermore, we investigated the approximate NLO PDF uncertainty in the $\rho_\mathrm{s}$ distribution using four different modern NLO PDF sets CT18NLO, ABMP16, MSHT20 and NNPDF3.1 and the dynamical scale $\mu_0=H_T^B/4$. 
In the bulk of the $\rho_\mathrm{s}$ distribution the predictions obtained with the different PDF sets agree well, while differences are observed in the high-energy tail, which are not covered by the PDF uncertainty. This was found to stem from differences in the large $x$-gluon distributions of the corresponding PDF fits. 

\section{Conclusions}
The $\rho_\mathrm{s}$ distribution seems to be better described, especially at low $\rho_{\mathrm{s}}$, when using the dynamical scale $H_T^B/4$ instead of the static scale $\mu_0=m_t$. This is concluded from the observation that the graphs obtained with different $(K_R, K_F)$ values do not cross using either described dynamical scale and the NLO and LO scale variation bands overlap over the whole $\rho_{\mathrm{s}}$ distribution, while using the static scale these only barely overlap in the high-energy region. The shape variation of the $\rho_\mathrm{s}$ distribution induced by the scale variation is smaller when adopting the dynamical scales instead of the static scale. This leads to a strongly reduced scale variation uncertainty in the normalized $\rho_\mathrm{s}$ distribution, that can be used for the experimental extraction of the top quark mass. In fact, using the dynamical scales the PDF uncertainty is comparable in size to the scale variation uncertainty in the normalized $\rho_s$ distribution. It was additionally found that while the size of the scale uncertainty does not show dependence on the $R$-parameter in the anti-$k_T$ jet clustering algorithm using either of the two dynamical scales, the statistics can be increased by using a larger $R$-value.

\bibliography{PROC_TOP2021}{}
\bibliographystyle{unsrt}
 
\end{document}

%% file: econfmacros.tex
\def\beq{\begin{equation}}
\def\eeq#1{\label{#1}\end{equation}}
\def\eeqn{\end{equation}}

\def\beqa{\begin{eqnarray}}
\def\eeqa#1{\label{#1}\end{eqnarray}}
\def\eeqan{\end{eqnarray}}

\let\bar=\overbar

\def\Dslash{\not{\hbox{\kern-4pt $D$}}}
\def\dslash{\not{\hbox{\kern-2pt $\del$}}}

\def\msb{{\bar{\ssstyle M \kern -1pt S}}}